\documentclass[psfig]{mn2e}
\usepackage{psfig}
\usepackage{times}

\def\gtrsim{\mathrel{\hbox{\rlap{\hbox{\lower3pt\hbox{$\sim$}}}\hbox{\raise2pt\hbox{$>$}}}}}
\def\lesssim{\mathrel{\hbox{\rlap{\hbox{\lower3pt\hbox{$\sim$}}}\hbox{\raise2pt\hbox{$<$}}}}}

\newcommand{\bd}{\begin{displaymath}}
\newcommand{\ed}{\end{displaymath}}
\newcommand{\be}{\begin{equation}}
\newcommand{\ee}{\end{equation}}

\newcommand{\msun}{{\rm M}_{\sun}}

\title{Warping of accretion disk and launching of jet by a spinning black hole in NGC 4258}

 \author[Wu et al.]
{Qingwen Wu\thanks{Corresponding author, E-mail: qwwu@hust.edu.cn}, Hao Yan, and Zhu Yi\\
    School of Physics, Huazhong University of Science and Technology, Wuhan 430074, China\\}

\pagerange{\pageref{firstpage}--\pageref{lastpage}}
\pubyear{}

\begin{document}

\maketitle
\label{firstpage}

\begin{abstract}
  We fit the most updated broadband spectral energy distribution from radio to X-rays for NGC 4258 with
  a coupled accretion-jet model that surrounding a Kerr black hole (BH), where both the jet and the
  warped $\rm H_2 O$ maser disk are assumed to be triggered by a spinning BH through
  Blandford-Znajek mechanism and Bardeen-Petterson effect respectively. The accretion flow
  consists with an inner radiatively inefficient accretion flow (RIAF) and an outer truncated
  standard thin disk, where the transition radius $R_{\rm tr}\simeq3\times 10^3R_{\rm g}$ for
  NGC 4258 based on the width and variability of its narrow Fe K$\alpha$ line. The hybrid jet
  formation model, as a variant of Blandford-Znajek model, is used to model the jet power.  Therefore,
  we can estimate the accretion rate and BH spin through the two observed quantities$-$X-ray emission and jet
  power, where the observed jet power is estimated from the low-frequency radio emission.
  Through this method, we find that the BH of NGC 4258 should be mildly spinning with dimensionless
  spin parameter $a_{*}\simeq0.7\pm0.2$. The outer thin disk mainly radiates at near infrared waveband and
  the jet contributes predominantly at radio waveband. Using above estimated BH spin and the
  inferred accretion rate at the region of the maser disk based on the physical existence of
  the $\rm H_2 O$ maser, we find that the warp radius is $\sim 8.6\times10^4 R_{\rm g}$ if it
  is driven by the Bardeen-Petterson effect, which is consistent with the observational result very well.

\end{abstract}


\section{Introduction}
   NGC 4258 (M106) is a low-luminosity, weakly active Seyfert 1.9 (Ho et al. 1997), located at a distance
   of $7.2$ Mpc (Herrnstein et al. 1999), where the active nucleus has been detected through radio (e.g,
   Turner \& Ho 1994), infrared (e.g, Chary et al. 2000) and X-ray observations (e.g, Fruscione et al. 2005;
    Reynolds et al. 2009). Wilkes et al. (1995) also detected the faint blue continuum emission and relatively
     broad ($\sim1000\rm\ km\ s^{-1}$, FWHM) emission lines in polarized light, which further support the
     existence of a weakly active nucleus. In the radio waveband, high-resolution interferometric images
     show the presence of a compact jet (e.g., Herrnstein et al. 1997), and two kpc-scale radio spots
     located at south and north of central engine respectively (e.g., Cecil et al. 2000). Furthermore,
     VLBI observations have found a set of $\rm H_2 O$ masers that trace a nearly edge-on, geometrically
     thin gas disk in the region of $\sim$ 0.14-0.28 pc from the nucleus, where the projected position of
      the maser spots indicates that the maser disk is substantially warped (Miyoshi et al. 1995). The nearly
      perfect Keplerian rotation curve of these $\rm H_2 O$ masers provides the precise measurement of the
       black hole (BH) mass $3.9\pm0.1\times10^7\msun$ and the inclination angle of $\sim 83^{\rm o}$ for
       the outer maser disk (Miyoshi et al. 1995; Herrnstein et al. 1999).

  The bolometric luminosity of NGC 4258 is small compared its Eddington luminosity
  ($L_{\rm bol}\sim 10^{-4}L_{\rm Edd}$, e.g, Reynolds et al. 2009, where $L_{\rm Edd}=1.3\times10^{38}(M_{\rm BH}/\msun)\rm erg\ s^{-1}$),
  which indicates that it is a typical low-luminosity active galactic nucleus (LLAGN). Its low luminosity
  may be caused by the radiative inefficiency of the accretion flow at low accretion rate, where the model
  of the hot, optically thin, geometrically thick radiatively inefficient accretion flow has been developed
   in last several decades (RIAF, an ¡®updated¡¯ version of the original advection dominated accretion flow
    model, ADAF; e.g., Ichimaru 1977; Naraya \& Yi 1994, 1995; see Narayan \& McClintock 2008 for a recent
    review).  The RIAF model can successfully explain most observational features of nearby LLAGNs (e.g.,
    Quataert et al. 1999; Cao \& Rawlings 2004; Nemmen et al. 2006; Wu et al. 2007; Yuan et al. 2009; Xu \& Cao 2009;
    see Ho 2008 for a recent review). For accretion rate less than a critical value (e.g., $\sim$1\% Eddington
    accretion rate), the standard thin disk (Shakura \& Sunyaev 1973) may transit to a RIAF at its inner region
    (e.g., Lu et al. 2004). The narrow Fe K$\alpha$ line and the ``red bump" that observed in LLAGNs do support
    this scenario (e.g., Quataert et al. 1999; Ho 2008; Xu \& Cao 2009; Shu et al. 2011). The spectrum of NGC 4258
    has been modeled with the truncated disk model, where the X-ray emission is explained by the original ADAF
    and the infrared data is fitted by the outer thin disk (Lasota et al. 1996; Gammie et al. 1999). Based on
    the possible power-law distribution of the infrared emission after correcting 18 mag of visual extinction,
    Yuan et al. (2002) proposed that the spectrum from infrared to X-rays can also be explained by a pure jet model.

   Although the warped disk of NGC 4258 has been observed nearly twenty years, the physical mechanism that
   responsible for producing such a structure is still unclear. There are several possible explanations for
   origin of the warped disk. Papaloizou et al. (1998) proposed that it could be produced by a binary companion
   orbiting outside the maser disk, where the companion object should has a mass comparable to that of disk.
   However, there is still no observational evidence for such an object. The second possibility is that the
   warped disk is caused by the non-axisymmetric forces due to the radiation pressure (Pringle 1996). However,
   this mechanism cannot produce the inferred warp in NGC 4258 if the accretion disk is radiatively inefficient
   (e.g., Maloney et al. 1996). For the possible radiatively efficient standard disk, it will be stable against
    the radiation instability and will not warp if the viscosity parameter of the maser disk $\alpha\lesssim0.2$
    (Caproni et al. 2007). Caproni et al. (2007) concluded that the warping of the maser disk in NGC 4258 should
    be caused by the Bardeen-Petterson effect, which is predicted in the framework of general relativity when the
    spin of a Kerr BH is inclined in relation to the angular momentum vector of the accretion disk (see also Martin 2008).
     The Lense-Thirring precession drives a warp in the disk, and the inner part of the disk is aligned with the BH
     (Bardeen \& Petterson 1975). There are some observational evidence that the Bardeen-Petterson effect may play
     a role in other compact objects (e.g., Liu et al. 2010; Lei et al. 2013).

   The jet formation mechanism has been extensively studied in the last several decades. The recent
   magnetohydrodynamic (MHD) simulations and analytical works found that both BH spin and large-scale magnetic
   field may play important roles in jet formation (e.g., McKinney \& Gammie 2004; Hirose et al. 2004; De Villiers
   et al. 2005; Hawley \& Krolik 2006; Hawley \& Krolik 2006; Li \& Cao 2009; Tchekhovskoy et al. 2011; Cao 2012;
   Cao \& Spruit 2013). In the Blandford-Znajek (BZ) process, energy and angular momentum are extracted from a
   spinning BH and transferred to a remote astrophysical load by open magnetic field lines. Narayan \& McClintock (2012)
   found that the scaling of the jet power and the BH spin in X-ray binaries is roughly consistent with that
   predicted by Blandford \& Znajek (1977), where the BH spins are determined via the continuum-fitting method
   (e.g., Zhang et al. 1997; Gou et al. 2011). Wu et al. (2011) also proposed that the radio FR I galaxies are
    powered by the fast rotating BHs based on the BZ process.

   NGC 4258 is the only AGN for which the BH mass, distance and the geometry of the outer disk are so accurately known.
   Both the multiwavelength observations of NGC 4258 and theories of the accretion flow have made important
    progress in last several years. For example, the X-ray observation of the narrow Fe K$\alpha$ line make it possible
     to estimate the transition radius of accretion flow (Reynolds et al. 2009); the strong variability of millimeter
     emission suggest it should originate in the region near the BH (Doi et al. 2013); the high-resolution near
     infrared data (Chary et al. 2000) and radio data from 1-43 GHz are also available (Doi et al. 2013), etc.
     These new updated data will help to further constrain the accretion and jet models compared former works
     (Lasota et al. 1996; Gammie et al. 1999; Yuan et al. 2002). On theoretical side, the global, time-dependent,
     numerical simulations revealed that the outflows may be important in RIAFs and only a fraction of the mass
     that available at large radii actually accretes onto the BH (e.g., Blandford \& Begelman 1999; Narayan et al. 2012;
     Yuan et al. 2012). Furthermore, in the old ADAF model, the gravitational energy is assumed to
     predominately heat the ions. However, it was later realized that the electrons may receive an important
     fraction of the dissipated energy through the magnetic reconnection and/or the viscosity arising from
     anisotropic pressure (e.g., Sharma et al. 2007). Therefore, it deserve to re-explore the central engine of
     this source with most updated multiwavelength observations and theories. The coupled accretion-jet model
     surrounding a Kerr BH is used in this work, which is not considered in former works. We further
     investigate the possible jet formation and the warped maser disk that may regulated by the spinning BH.

   For a distance of 7.2 Mpc and a BH mass of $3.9\times10^7\msun$  for NGC 4258, $1''$ corresponds to 35 pc
   and 1 pc corresponds to $5.4\times10^5 R_{\rm g}$, where the gravitational radius $R_{\rm g}=GM_{\rm BH}/c^2$.

\begin{table*}
 \begin{flushleft}
   \centering
 \centerline{\bf Table 1. Spectral energy distribution for NGC 4258.}
  \begin{tabular}{lllllc}\hline
Waveband  & $\nu$                    & Flux  & Luminosity        & Resolution    & References \\
           &  Hz                      & mJy   & $\rm erg\ s^{-1}$ &               &  \\
\hline
1.4 GHz    & 1.4$\times10^{9}$    &   2.73          &  35.37            &$\sim1^{''}$           & 1   \\
5   GHz    & 5$\times10^{9}$      &   2.0$\pm0.1$   &  35.79$^{+0.02}_{-0.02}$  &$\sim0.5^{''}$ & 2   \\
8.4 GHz    & 8.4$\times10^{9}$    &   2.2$\pm0.1$   &  36.06$^{+0.02}_{-0.02}$  &$\sim0.5^{''}$ & 2   \\
15  GHz    & 1.5$\times10^{10}$   &   2.7$\pm0.2$   &  36.40$^{+0.03}_{-0.03}$  &$\sim0.5^{''}$ & 2   \\
22  GHz    & 2.2$\times10^{10}$   &   2.9$\pm0.4$   &  36.65$^{+0.05}_{-0.07}$  &$\sim0.5^{''}$ & 2   \\
22  GHz    & 2.2$\times10^{10}$   &   $<$0.22       &  $<35.48^{+0.3}$       &$\sim0.0003^{''}$ & 3   \\
43  GHz    & 4.3$\times10^{10}$   &   6.9$\pm0.2$   &  37.26$^{+0.01}_{-0.01}$  &$\sim0.1^{''}$ & 2   \\
100 GHz    & 1$\times10^{11}$     &   6.9$\pm0.2$   &  37.63$^{+0.01}_{-0.01}$  &$\sim10^{''}$  & 2   \\
347 GHz    & 3.47$\times10^{11}$  &   93.7$\pm20.8$ &  39.30$^{+0.09}_{-0.11}$  &$\sim20^{''}$  & 2   \\
17.9$\mu$m & 1.68$\times10^{13}$  &   300$\pm30$    &  41.49$^{+0.05}_{-0.06}$  &$\sim0.5^{''}$ & 4   \\
12.5$\mu$m & 2.40$\times10^{13}$  &   165$\pm20$    &  41.39$^{+0.06}_{-0.06}$  &$\sim0.5^{''}$ & 4   \\
10.5$\mu$m & 2.86$\times10^{13}$  &   100$\pm12$    &  41.25$^{+0.05}_{-0.06}$  &$\sim6^{''}$   & 4   \\
3.45$\mu$m & 8.70$\times10^{13}$  &   20 $\pm3$     &  41.03$^{+0.06}_{-0.07}$  &$\sim0.2^{''}$ & 4   \\
2.2$\mu$m  & 1.36$\times10^{14}$  &   4.0$\pm0.7$   &  40.53$^{+0.08}_{-0.12}$  &$\sim0.2^{''}$ & 4   \\
1.6$\mu$m  & 1.88$\times10^{14}$  &   0.9$\pm0.3$   &  40.02$^{+0.12}_{-0.18}$  &$\sim0.2^{''}$ & 4   \\
1.1$\mu$m  & 2.73$\times10^{14}$  &   $<$0.5        &  $<$39.93                 &$\sim0.2^{''}$ & 4   \\
$B$-band   & 6.81$\times10^{14}$  &   ...           &  39.97$^{+0.30}_{-0.29}$  &...            & 5   \\
2-10 keV   & ...                  &   ...           &  $40.94^{+0.16}_{-0.20}$  &...            & 6   \\
\hline
\end{tabular}
\end{flushleft}
\end{table*}
\begin{table*}
\begin{minipage}{170mm}
References: (1) Ho \& Ulvestad (2001); (2) Doi et al. (2013); (3) Herrnstein et al. (1998); (4) Chary et
al. (2000); (5) this work; (6) Fruscione et al. (2005).
 \end{minipage}
\end{table*}

\section{The nuclear SED}
   The radio emission of NGC 4258 has been thoroughly investigated from $\sim$ 1 GHz upto millimeter waveband (see
   Doi et al. 2013, and references therein). We simply summarize its radio properties as follows. A two-sided nuclear
   jet extending up to $\sim 10^6 R_{\rm g}$ was found in radio observation at $\sim$1.5 GHz (Cecil et al. 2000),
   which may physically connect with the kpc-scale jet-like structures (¡°anomalous arms¡±, e.g., van der Kruit et al.
   1972). Herrnstein et al. (1998) reported a jet-like structure in 22 GHz radio observation, which oriented along
   the rotation axis of the maser disk, where the northern and southern components have flux density of 2.5-3.5 mJy
   and 0.5 mJy respectively. Interestingly, the VLBI images also provided a 3$\sigma$ upper limit of 220$\mu$Jy on
   22 GHz at the dynamical center, which place a strong constraint on accretion disk (Herrnstein et al. 1998), where
    the actual upper limit is likely 1-2 times higher than the observed 220$\mu$Jy if considering the possible
    free-free absorption in the disk (e.g., Herrnstein et al. 1996). Other multi-frequency radio observations
    (e.g., 5, 8.4, 15 and 43 GHz) at $\sim0.5^{''}$ resolution toward the nucleus are also presented in Nagar et al.
    (2001) and Doi et al. (2013). Doi et al. (2013) presented the millimeter and submillimeter observations for NGC 4258,
     where the strong variability at 100 GHz observed by the Nobeyama Millimeter Array (NMA) suggest it should come
     from the region close the BH, while most of the submillimeter emission at 347 GHz observed by James Clerk Maxwell
     Telescope (JCMT) may originate in interstellar dust of the host galaxy. In building the radio spectrum, the
      average flux is used if there exist multi-epoch data at given waveband (see Table 1 and Doi et al. 2013, for details).

   Chary et al. (2000) presented high-resolution infrared imaging of the nucleus in NGC 4258 from 1 to 18 $\mu$m.
   The compact source of radiation is still unresolved even at near-infrared resolution of $0.2^{''}$. Chary et al.
   (2000) found that the near- and mid-infrared data can be fitted by a single power law as $\nu^{-1.4\pm0.1}$ if
   the infrared data was corrected for about 18 mag of visual extinction, and they suggest that the power-law
   infrared data may originate from the nonthermal emission. However, Mason et al. (2012) found that the mid-infrared
   emission of NGC 4258 is similar to those of conventional bright Seyferts, and the quite evident silicate emission
    features demonstrate that the dust should present in its nuclear region. Therefore, the possible power-law
    distribution of infrared emission as proposed by Chary et al. (2000) may be not intrinsic. The $observed$
    high-resolution infrared data are listed in Table 1 (see Chary et al. 2000 for more details).

   The optical emission of NGC 4258 is less clear, due to the possible obscuration by the putative torus
   or/and warped disk. The indirect estimation of the optical luminosity is $L_{5500\rm \AA}\sim10^{37}-10^{39}\rm erg\ s^{-1}\AA^{-1}$
   from the nuclear polarized light (Wilkes et al. 1995). However, this estimation still suffers many uncertainties. In this work,
    we estimate the nuclear optical emission of NGC 4258 from its H$\beta$-line luminosity, using the correlation between
    $L_{\rm H\beta}$ and $M_{B}$ for both bright and faint type I AGNs, as calibrated by Ho \& Peng (2001).
    From this method, its $B$-band luminosity is $L_{B}=10^{39.97^{+0.30}_{-0.29}}\ \rm erg\ s^{-1}$, where
    the distance-corrected $\rm H\beta$-line luminosity $L_{\rm H\beta}=10^{38.82}\ \rm erg\ s^{-1}$ and the
    empirical correlation $\log L_{\rm H\beta}=(-0.34\pm0.01)M_{B}+(35.1\pm0.25)$ were adopted (Ho \& Peng 2001).

   The X-ray emission of NGC 4258 is known to be variable on timescale of hours to years with unabsorbed
   2-10 keV X-ray luminosity of (4.2-17.4)$\times10^{40}\ \rm erg\ s^{-1}$, where the average X-ray
   luminosity is $8.8\times10^{40}\rm erg\ s^{-1}$ and the average photon index is $1.6\pm0.2$ (see
    Fruscione et al. 2005). Using the data of $Suzaku$, Reynolds et al. (2009) proposed that the region
    of the hard X-ray emission should be smaller than 25$R_{\rm g}$ based on the high-amplitude intraday
    variability. They further constrained that the narrow Fe K$\alpha$ line emitting region to be ranged
    from $\sim 3\times10^3 R_{\rm g}$ to $\sim 4\times10^4 R_{\rm g}$ based on the width and flux
    variability of Fe K$\alpha$ line by comparing the $Suzaku$ data with $XMM$-$Newton$ data that taken 160 days later.

   All the data for building the SED of NGC 4258 and related references are listed in Table 1.

\section{Model for the central engine}

  Our model consists three components: an inner RIAF, an outer truncated thin disk that may
  connect to the warped maser disk and a jet, which are illustrated in Figure 1. We describe
  these components in details as follows.

  \subsection{The inner RIAF}
  The RIAF model surrounding a Kerr BH is considered in this work. Only main features of the
  model are described here, and the details of which can be found in Wu \& Cao (2008, and references
   therein). We employ the approach suggested by Manmoto (2000) to calculate the global structure of
   the RIAF in the general relativistic frame, which allows us to calculate the structure of a RIAF
   surrounding either a spinning or a nonspinning BH. Both observations and theories suggest the
   existence of outflow in hot accretion flows. To describe the possible mass loss of the disk, we
   assume that the accretion rate is a function of radius, $\dot{M}=\dot{M}_{\rm tr}(R/R_{\rm tr})^{\it s}$,
   where $R_{\rm tr}$ is the transition radius of the disk and $\dot{M}_{\rm tr}$ is the accretion rate at
   $R_{\rm tr}$ (e.g., Blandford \& Begelman 1999). Yuan et al. (2012) found that the profile of $\dot{M}(R)$
   has two parts from the simulations. When $R\lesssim10R_{\rm g}$, $s\sim0$; but outside $10R_{\rm g}$, $s\sim0.5$.
   In the present paper, we set $s$ = 0.4 throughout the flow for simplicity, which is also roughly consistent
    with that constrained from the observation of the supermassive BH in our Galactic center-Sgr A*
    (Yuan et al. 2003). The global structure of the RIAF surrounding a BH that spinning at a rate $a_{*}$ with
     mass $M_{\rm BH}$ can be calculated with proper outer boundaries, if the parameters $\dot{m}$, $\alpha$,
     $\beta$ and $\delta$ are specified (e.g., Manmoto 2000), where all radiation processes (Synchrotron,
     Bremsstrahlung and Compton scattering) are included self consistently. The parameter $a_{*}=Jc/GM_{\rm BH}^{2}$
     is the dimensionless angular momentum of the BH ($J$ is the angular momentum of the BH), and
     $\dot{m}=\dot{M}/\dot{M}_{\rm Edd}$ is the dimensionless accretion rate (the Eddington accretion
     rate is defined as $\dot{M}_{\rm Edd}=1.3\times10^{18}M_{\rm BH}/\msun \rm\ g \ s^{-1}$). For the
     viscosity parameter $\alpha$, we adopt the typical value of $\alpha=0.2$ which is constrained from
     the dwarf novae in hot state or the modeling the spectrum of LLAGNs ($\alpha\simeq0.1-0.3$, Narayan
     \& McClintock 2008, and references therein). Numerical simulations suggest that magnetic fields are
     generally subthermal, with $P_{\rm mag}/P_{\rm gas}\sim0.1$ (Hawley et al. 2006), so we expect the
      ratio of the gas to the total pressure (sum of gas and magnetic pressure) $\beta=P_{\rm g}/P_{\rm tot}=0.9$.
      The parameter $\delta$ describes the fraction of the turbulent dissipation that directly heats the
      electrons in the flow. In this work, we adopt $\delta=0.1$, which can well describe the
      multiwavelength radiation of LLAGNs (Liu \& Wu 2013). The field-enhancing effect caused by
      frame dragging is also considered (Meier 2001). The amplified magnetic field related to the
      magnetic field produced by the dynamo process in the RIAF can be expressed as $B=gB_{\rm dynamo}$,
      where $g=\Omega/\Omega^{'}$ is the field enhancing factor, the disk angular velocity $\Omega$ is a
      sum of its angular velocity relative to the local metric $\Omega^{'}$ plus the angular velocity of
      the metric itself in the Boyer-Lindquist frame $\omega\equiv-g_{\phi t}/g_{\phi\phi }$, i.e.,
      $\Omega=\Omega^{'}+\omega$. The spectrum is calculated using the global structure of RIAF surrounding
      a Kerr BH, where the gravitational redshift effect is considered, while the relativistic optics near
      the BH is neglected (e.g., Manmoto 2000).

 \begin{figure}
  \centerline{\psfig{figure=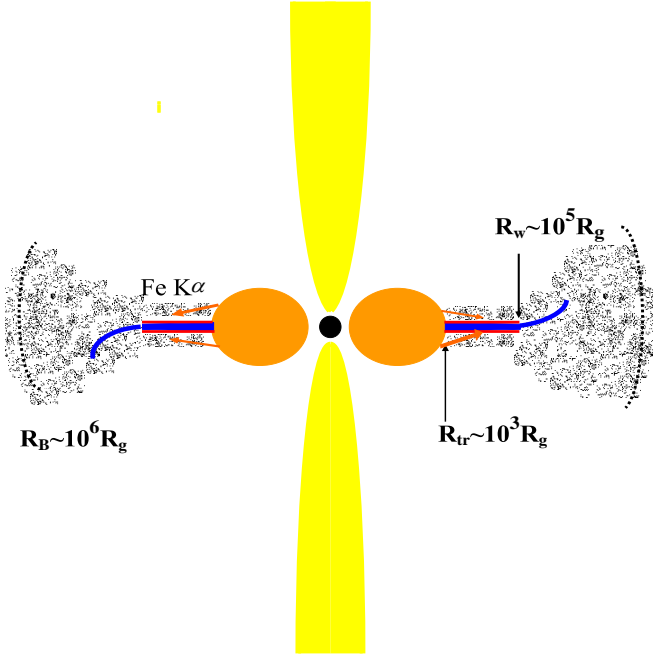,width=9cm,height=9cm}}
  \caption{The cartoon illustrating the model for the central engine of NGC 4258. It mainly
  consists of three components: an inner RIAF, an outer truncated thin disk which may connect
  to outer warped maser disk, and a jet. The dotted lines represent the Bondi radius, and the
  hot gas inside this radius will also be captured by BH. } \label{f1}
  \end{figure}

 \subsection{The outer thin disk}
   Outside the inner RIAF there is an outer thin disk with a truncation radius of $R_{\rm tr}$.
   The physical mechanism for the disk transition is still an open issue. In this work, we set
   $R_{\rm tr}=3000 R_{\rm g}$ for NGC 4258, which is constrained from the  width and flux
   variability of the narrow Fe K$\alpha$ line (Reynolds et al. 2009). The outer radius of
   the thin disk is unclear, which may smoothly connect to the observed maser disk.

   In addition to the accreting cold gas through the maser disk at large scale, the hot
   plasma around the Bondi radius also provides a suitable mass supply rate for BH accretion
   (e.g., M 87, Di Matteo et al. 2003). The Bondi radius $R_{\rm B}=GM_{\rm BH}/c_{\rm s}^2\simeq 10^6R_{\rm g}$
   for a typical temperature of 1 keV for the diffused hot gas at $\sim R_{\rm B}$, where
   $c_{\rm s}\simeq0.1 T^{1/2}\rm km\ s^{-1}$ is the sound speed of the hot gas. The Bondi
   radius is around one order of magnitude larger than the observed outer radius of maser
   disk ($\sim0.28\rm pc\sim 1.5\times 10^5R_{\rm g}$). Therefore, the inflow initially should
   consist a two-phase accretion flow where the cold thin disk is sandwiched by hot plasma. The
   hot inflow inside the Bondi radius will be gradually cooled down and condensed into the thin
   disk due to the increase of the radiative efficiency through the inverse Compton process when
   the cold disk is present. Based on this scenario, we think that it is possible for the
   accretion rate at truncation radius, $\dot{M}_{\rm tr}$, larger than that of the maser disk,
    $\dot{M}_{\rm m}$. Most of the radiation from the truncated thin disk originate in the
    region near the transition radius due to the $1/R$ dependence of the gravitational potential.
     Therefore, we calculate the spectrum of outer thin disk simply using a constant rate at $R_{\rm tr}$,
     which will not affect our main conclusion. The thin disk emits locally as a blackbody, and its
     spectrum can be calculated if BH mass, accretion rate and inclination angle of the disk are known.
     We adopt the inclination angle of the inner disk to be $74^{\rm o}$ that may aligned with the
     equatorial plane of the BH (e.g., Krause et al. 2007).

 \subsection{The jet: power \& spectrum}

  We investigate both the total power and radiative spectrum of the jet in NGC 4258, since that the
  radio jet has been detected directly. To evaluate the jet power extracted from the inner region of
  RIAF that surrounding a spinning BH, we adopt the hybrid model as a variant of a BZ model  proposed
  by Meier (2001). The total jet power is given by
  \be
  P_{\rm jet} = B_{\rm p}^{2}R^{4}\Omega^{2}/32c,
  \ee
  where $R$ is the characteristic size of jet formation region, the poloidal magnetic field
  $B_{\rm p}\simeq gB_{\rm dynamo}$ (see Section 3.1 and Meier 2001 for more details). Following
  the work by Nemmen et al. (2007), all the quantities are evaluated at the innermost marginally
  stable orbit of the disk $R=R_{\rm ms}$, which is roughly consistent with the jet launching region
  in the numerical simulations. Wu et al. (2011) found that the jet efficiency of this model is
  consistent with those of MHD simulations very well.

  Both the protons and electrons will be accelerated by the magnetic field energy that extracted
  from the disk and/or rotating BH, and the kinetic energy of particles will be eventually
  released through the possible shock in the jet. To describe the jet radiation, we adopt the
  internal shock scenario that has been used to explain the broadband SED of X-ray binaries,
  AGNs and afterglow of gamma-ray burst (e.g., Piran 1999; Yuan et al. 2005; Wu et al. 2007;
  Nemmen et al. 2012). The jet is assumed to be perpendicular to the underlying RIAF
  and have a conical geometry with opening angle $\phi$. The internal shock in the jet occur
  as a result of the collision of shells with different velocities, and these shocks accelerate
  a small fraction, $\xi_{\rm e}$, of the electrons into a power-law energy distribution with
  index $p$. From the shock jump conditions, the electron and ion temperatures can be derived.
  Following the widely adopted approach in the study of GRBs, the energy density of accelerated
  electrons and the amplified magnetic field are described by two free parameters
  $\epsilon_{\rm e}$ and $\epsilon_{\rm B}$, which roughly follow $\epsilon_{\rm e}=\epsilon_{\rm B}^{1/2}$
  (Medvedev 2006). We then calculate the radiative transfer by both thermal and power-law electrons in the
  jet, although we find that the latter plays a dominant role. Only synchrotron emission is considered,
  since Compton scattering is not important in this case (e.g., Markoff et al. 2001; Wu et al. 2007).

  The 22 GHz radio observations show that NGC 4258 have a broad jet, where the jet open angle
  $\sim 20^{\rm o}$ (Herrnstein et al. 1997), which is similar to the jet in other nearby LLAGNs
  (e.g., M81, Markoff et al. 2008), and hence we adopt $\phi=20^{\rm o}$. The typical value of
   $\epsilon_{\rm e}$=0.1 for LLAGNs is adopted in this work (e.g., Yuan et al. 2009; Nemmen et al. 2012),
   and then $\epsilon_{\rm B}$=0.01 according to the relation proposed by Medvedev (2006). Shock
   acceleration theory typically gives $2\lesssim p \lesssim 3$, and we fix $p=2.5$. For the fraction
   of electrons that were accelerated into the power-law distribution, we use $\xi_{\rm e}=0.1$ as
   that of FR I radio galaxies (Wu et al. 2007). There are two free parameters in the jet model:
   the outflow rate $\dot{m}_{\rm jet}$ and the jet velocity $v_{\rm jet}$, and these two parameters
   can be determined by fitting the spectrum and the jet power that estimated from equation (1).

\section{Result}

 \subsection{Estimating the jet power}
  We estimate the jet power of NGC 4258 from its low-frequency radio emission since
  it usually originates in the diffused optically thin lobes that move at low velocities
  and do not suffer the Dopper boosting effect. Wu et al. (2011) derived an empirical
  correlation between 151 MHz radio luminosity and jet power that measured from X-ray
  cavities for a sample of low-power radio galaxies, which is $\log P_{\rm jet}=0.52\pm0.04\log L_{151 \rm MHz}+22.31\pm 1.75$
  with an intrinsic scatter of 0.23 dex. Through this correlation and 151 MHz radio luminosity of
  $37.28\ \rm erg\ s^{-1}$ reported in Hales et al. (1988), we find that jet power of NGC 4258 is
  $10^{41.69}\ \rm erg\ s^{-1}$. The estimated jet power is consistent with that reported in
   Falcke \& Biermann (1999) very well, where the jet power of $10^{41.7}\ \rm erg\ s^{-1}$ is
   derived from the 22 GHz nuclear radio emission and offset of the core from the dynamical center.

\subsection{Modeling the multiwavelength spectrum}
   The SED of NGC 4258 is plotted in Figure 2. In spectral fitting, we, firstly, adjust the
   parameters of BH spin, $a_*$, and the accretion rate at the transition radius, $\dot{m}_{\rm tr}$,
   to fit the observed X-ray spectrum and simultaneously allow the derived jet power from equation
   (1) to be consistent with the above estimated jet power, since that both jet power and high-energy
   X-ray emission are predominantly originate from the inner region of the RIAF near the BH. We find
   that the best-fit parameters are $a_*\simeq0.71$ and $\dot{m}_{\rm
   tr}\simeq8\times10^{-3}$. Considering the possible uncertainties in estimation of jet power (0.23
   dex), the BH spin parameter is $a_*\simeq0.71^{+0.12}_{-0.18}$. It is interesting
   to note that the emitted spectrum from the underlying RIAF (dashed line) can not only reproduce the
   observed X-ray spectrum fairly well; but also roughly consistent with the estimated nuclear optical
   emission and the upper limit of 22 GHz emission that radiated from the central engine (red solid and
   open circles in Figure 2). Secondly, we show the emitted spectrum from the outer thin disk in Figure 2
   (dot-dashed line), where the transition radius of $R_{\rm tr}=3000 R_{\rm g}$ and the accretion rate
   $\dot{m}_{\rm tr}$ are used directly in spectral calculations. We find that the outer SSD mainly
   radiates at near infrared waveband. Thirdly, the radio emission of NGC 4258 should originate in
   the jet since it was observed directly. We find that the jet model with outflow rate,
   $\dot{m}_{\rm jet}=5\times10^{-6}$ and jet velocity, $v_{\rm jet}=0.86c$, can well reproduce
   both the observed radio spectrum (dotted line) and the estimated jet power, where $\sim 0.83\%$
   of the mass that ultimately accreted by the black hole is channeled into the jet. The solid
   line in Figure 2 is the sum of the spectrum from the inner ADAF, outer thin disk and the jet.
   We find that our model gives a quite nice fit for the observed spectrum except the mid-infrared emission.

\begin{figure}
  \centerline{\psfig{figure=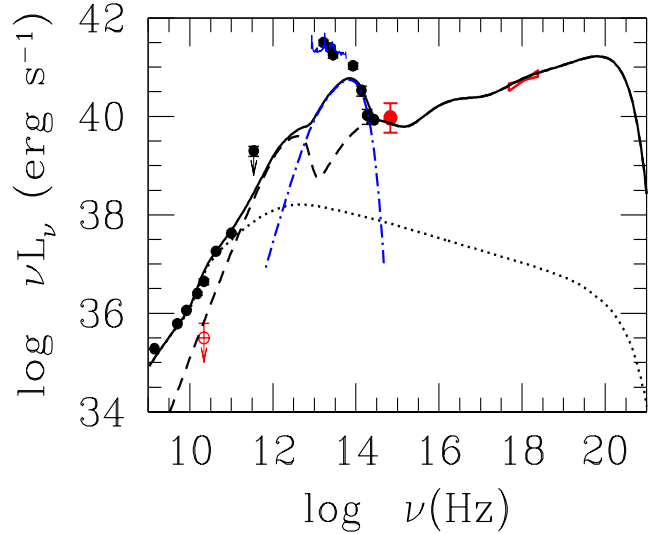,width=9.cm,height=7cm}}
  \caption{The spectral energy distribution of NGC 4258. The dashed, dot-dashed and dotted lines
  show respectively the emission from the inner RIAF, outer SSD and jet, while the solid line
  shows the sum of the radiation from these components. The red solid circle represent the nuclear
  optical emission estimated from the H$\beta$-line luminosity, and the red open circle represent
  the upper limit of the 22 GHz radio emission that coincident with the central engine. The blue
  solid line denote the MIR spectrum of NGC 4258 that chosen from Mason et al. (2012). } \label{f2}
  \end{figure}

\subsection{The Bardeen-Petterson effect and warp radius of outer maser disk}
  Frame dragging produced by a Kerr BH causes precession of a particle if its orbital plane
  is inclined in relation to the equatorial plane of the BH, which is known as Lense-Thirring
  precession. There exists a critical radius, $R_{\rm w}$, in the warped disk, where the disk
  material interior to $R_{\rm w}$ will be aligned with the equatorial plane of the black hole,
  while the outer portion of the disk ($R>R_{\rm w}$) will maintain its original orientation
  (Bardeen \& Petterson 1975). The critical warp radius $R_{\rm w}$ can be determined
  by assuming the angular momentum of the spinning BH, $J_{\rm BH}=a_{*}GM^2_{\rm BH}/c$, to
   be equal to the angular momentum of the disk in per logarithmic interval of the radius,
   $J(R)=2\pi R^3\Sigma v_{\phi}$, where $\Sigma$ is the surface density of the disk and
   $v_{\phi}$ is the rotational velocity of the disk material (Sarazin et al. 1980; Lu \& Zhou 2005).
   Based on the standard thin disk model, the critical radius is (Lu \& Zhou 2005),
  \be
   \frac{R_{\rm w}}{R_{\rm g}}=7.1\times10^3\alpha^{16/35}a_{*}^{4/7}\left(\frac{M_{\rm BH}}{3.9\times10^7\msun}\right)^{5/7}
    \left(\frac{\dot{M}}{\dot{M}_{\rm Edd}}\right)^{-2/5}.
  \ee
  The motion of the infalling gas of NGC 4258 cannot be measured directly, and there are some indirect methods
   to estimate the accretion rate in the \emph{maser} disk. The maser amplification is favored when the particle
    density range from $10^8$ to $10^{10}\ \rm cm^{-3}$ (Moran et al. 1995). Cao \& Jiang (1997) derived an
    accretion rate of $\dot{M}_{\rm m}\simeq8\times10^{-4}\alpha \msun\rm yr^{-1}$ at the inner edge of
    the maser disk in NGC 4258 based on the standard thin disk model, where they assume the disappearance
   of the maser disk at the inner region is caused by the density of the molecular gas in disk reach the
   upper limit for the existence of the maser. Using above estimated BH spin $a_*\simeq0.71^{+0.12}_{-0.18}$ and accretion
   rate in maser disk, we find that the warp radius $R_{\rm w}\simeq1.0^{+0.09}_{-0.15}\times10^5 \alpha^{2/35} R_{\rm g}$.
   For the typical value of $\alpha\sim0.1$, the warp radius $R_{\rm w}\sim8.6\times10^4 R_{\rm g}$, which
   is evidently not sensitive to the viscosity parameter.

\section{Discussion}

   The observational features of NGC 4258, such as the hard X-ray spectrum, narrow Fe $\rm K \alpha$
   line, infrared bump, and the variable inverted radio spectrum, indicate that it is most likely a
   typical LLAGN (Ho 2008).  NGC 4258 is also one of the nearest LLAGNs that suitable for studying
   the properties of their central engines, following the extremely sub-Eddington accreting source-Sgr A*
   at the Galactic center (8 kpc, Reid 1993) and the closest LLAGN-M81 (3.6 Mpc, Freedman et al. 2001).
   The warped maser disk at sub-parsec scales in NGC 4258 places strong constraints on both its BH mass,
   disk geometry and even the possible accretion rate in the maser disk. The inner RIAF + outer thin
   SSD + jet have been considered as the standard model for nearby LLAGNs and low/hard state X-ray
   binaries (e.g., Quataert et al. 1999; Nemmen et al. 2006; Xu \& Cao 2009, and see Ho 2008 for a
   recent review), where the inner RIAF mainly radiates at hard X-ray band, the outer SSD radiates
   at infrared to optical band, while the jet mainly radiates at radio waveband. Our detailed modeling
   of NGC 4258 do support this scenario.  We investigate the properties of the central engine in NGC 4258
   with a accretion-jet model that surrounding a Kerr BH, where we find that a mildly rotating BH with
   $a_*\simeq0.71$ can well reproduce the hard X-ray emission, jet power and warp radius of the maser disk.

    We estimate the BH spin of NGC 4258 based on the accretion-jet model, where the accretion flow
    is described by the inner RIAF + outer SSD. There are some degeneracies of the model parameters in
    accretion-jet model. For example, the transition radius $R_{\rm tr}$ is degenerated with the accretion
    rate $\dot{m}_{\rm tr}$ in fitting the near IR spectrum; the parameters $s$, $\dot{m}_{\rm tr}$,
    $R_{\rm tr}$ and $\delta$ are also degenerated for reproducing the observed X-ray spectrum. It should
     be noted that both the X-ray emission and jet power predominantly come from the inner region of the
     accretion flow near the BH (e.g., $\lesssim 10R_{\rm g}$), where most of the gravitational energy of
     the accreting material released. Therefore, both the X-ray emission and jet power are only sensitive
     to the accretion rate at the inner edge of the accretion flow $\dot{m}_{\rm in}$, BH spin and the
     parameter $\delta$, where the parameter $\delta$ describes the fraction of the turbulent dissipation
     that directly heats the electrons in accretion flow. We can estimate these two parameters $a_{*}$
     and $\dot{m}_{\rm in}$ using the observed X-ray emission and jet power for given $\delta$.
     For estimation of the jet power, we use the empirical correlation between  $L_{\rm 151\ MHz}$
     and $P_{\rm jet}$ that derived from a sample of the radio-filled sources of which X-ray cavities have been observed
     (Wu et al. 2011), and the estimated jet power for NGC 4258 in this method is quite consistent
      with that derived from other method (e.g., Falcke \& Biermann 1999). The scatter of the adopted relation ($\sim0.23$ dex)
     is much smaller than those of Birzan et al. (2008, $\sim 0.8$ dex) and Cavagnolo et al. (2010, $\sim0.7$
     dex), where the ghost cavities are included. Considering
     the uncertainty of 0.23 dex for estimation of the jet power and a typical value of $\delta=0.1$,
     we find that $a_{*}\simeq0.71^{+0.12}_{-0.18}$ and $\dot{m}_{\rm in}\simeq4.3\times10^{-4}$ for NGC 4258.
     We note that the degeneracy of the parameters $s$, $\dot{m}_{\rm tr}$ and $R_{\rm tr}$ will not
     affect the estimation of BH spin provided the parameters $\dot{m}_{\rm in}$ and $\delta$ are the
     same, where $\dot{m}_{\rm in}$ is regulated by the degenerated parameters $-s, \dot{m}_{\rm tr}, R_{\rm tr}$.
     Sharma et al. (2007) found that the parameter $\delta$ may be in range of $\sim0.03-0.3$ based on the
     simulations, depending on the model details. To reproduce the observed X-ray luminosity and the estimated
     jet power, we find that the spin should be \textbf{$a_*\simeq 0.85^{+0.05}_{-0.09}$ }for $\delta=0.3$,
     where $\dot{m}_{\rm in}\simeq1.1\times10^{-4}$
     becomes slightly lower due to more dissipated energy is directly heated electrons and, therefore, the
    lower accretion rate will be needed to reproduce the X-ray spectrum. In this case, the higher BH spin
     is needed to reproduce the observed jet power. For the case of lower value of $\delta=0.03$, the
    slightly higher $\dot{m}_{\rm in}\simeq5.34\times10^{-4}$ and lower BH spin with $a_*\simeq 0.63^{+0.14}_{-0.17}$
    are needed to reproduce the observed X-ray emission and jet power. Therefore, the BH of NGC 4258 may be moderate
    spinning with $a_*\simeq 0.7\pm0.2$ considering the possible uncertainties
    in $\delta$ parameter and the estimation of jet power. Even the adopted hybrid
    jet model based on the RIAF is quite consistent with some numerical magnetohydrodynamic(MHD)
    simulations (see Figure 1 in Wu et al. 2011), it should be kept in mind that there are still
    some uncertainties for the jet formation that regulated by a spinning BH (e.g.,
    Martinez-Sansigre \& Rawlings 2011, and references therein). For example, the jet efficiency
    could be close 100\% or even higher according to recent general relativistic MHD simulation based on
    the magnetically arrested thick accretion flow ($H/R\sim 1$, e.g., McKinney et al. 2012). If this is the case,
    we may overestimated the BH spin, even it is still unclear whether the accretion flow of NGC 4258
    has so strong magnetic field or not.
    Furthermore, the jet power may has a steeper dependence on the spin for the case of a thick disk with
   $H/R\gtrsim1$ (e.g., McKinney 2005, Tchekhovskoy et al. 2010). But the recent
   observational results from the black-hole X-ray binaries seem to be still support the transitional BZ
   mechanism ($P_{\rm jet}\propto a_{*}^2$, Narayan \& McClintock 2012).

    We adopt the shock scenario of the jet model to fit the radio emission of NGC 4258, since the radio
    emission is emitted predominantly by the jet. In spectral fitting, we find that the ratio of the mass
    loss rate in the jet to the accretion rate near the horizon of the BH is $\sim1\%$, which is a typical
    ratio for BH that accreting at a low rate (Yuan \& Cui 2005; Yu et al. 2011; Nemmen et al. 2012). Using
    this outflow rate, we find that the speed of the jet is $\sim 0.86 c$, which is mildly relativistic.
    Considering the possible slightly lower fraction of electrons are accelerated into the power-law
    distribution, e.g., $\xi_{\rm e}=0.01$, in less powerful jet of NGC 4258 compared that of $\xi_{\rm e}=0.1$
    in the powerful jets of FR Is, we find that the ratio of the mass outflow rate will be slightly higher,
    $\sim4.6\%$, and the outflow speed will be lower, $v_{\rm jet}=0.6 c$, for given estimated jet power.
    Therefore, the jet in NGC 4258 should be mildly relativistic even considering possible uncertainties
    in model parameters. In fact, the mildly relativistic jets have been inferred from observations of
    other LLAGNs/nearby Seyfert galaxies, e.g., Arp 102B (0.45c; Fathi et al. 2011) and NGC 4278 (0.76c;
     Giroletti et al. 2005). Most importantly, we find that the jet model cannot reproduce the radio and
     infrared/X-ray emission simultaneously, even we allow all parameters of jet model to be free.
      Therefore, the nuclear SED of NGC 4258 is most probably produced by the coupled accretion and jet.

    The 6.4 keV Fe K$\alpha$ line (e.g., width, strength, and variability) is one of the most powerful
    probes for the cold gas surrounding the central BH. Assuming the Keplerian orbits in the outer SSD,
    Reynolds et al. (2009) estimated the line emitting region roughly range from $3\times 10^3$ to
    $4\times 10^4R_{\rm g}$ using the line width and variability, which suggest that the outer SSD
    may transit to an ADAF at $R_{\rm tr}\lesssim 3\times 10^3 R_{g}$. If this is the case, we find
    that the accretion rate $\dot{m}_{\rm tr}=8.0\times10^{-3}$ can well reproduce the observed near
    infrared data. The mid-infrared emission ($\gtrsim 10\mu\rm m$) is much higher than that predicted
    by the outer SSD, where the origin of the MIR emission is still unclear and it may come from other
    components. The slightly lower resolution of mid-infrared data with $0.5''-20''$ are easily
     suffered the contamination from the host galaxies. Laine et al. (2010) found that the mid-infrared
     emission of NGC 4258 at 8$\mu\rm m$ that obtained from the $Spitzer$ is mainly originate from the
     polycyclic aromatic hydrocarbon molecules and/or hot dust. In the model fitting, we have assumed
     that the NIR emission is come from the truncated outer SSD. Therefore, the confirmation of the
     NIR origin should be crucial to our work. The infrared variability may help to test this issue,
      where the NIR variability should correlated with the X-ray variability if it is dominated by
      the reprocessing of the putative torus while the emission should be steady over much longer t
      imescales if it come from the outer truncated thin disk. We note that these uncertainties
      will not affect our main conclusion on the estimation of BH spin, which is mainly controlled
    by the accretion-jet properties near the BH.

   The accretion rate in outer maser disk has been explored with different methods, which
    is $\sim (1-8)\times10^{-4}\alpha \msun/\rm yr$, where the rate $\sim 10^{-4}\alpha\msun/\rm yr$
    is estimated by assuming a transition from a molecular to an atomic disk at outmost region of
    maser disk (Neufeld \& Maloney 1995; Herrnstein et al. 2005); and the rate of $\sim 8\times 10^{-4}\alpha\msun/\rm yr$
    is estimated by assuming the quenched maser at its inner region is caused by the molecular density
    reach its upper limit of $10^{10}\rm cm^{-3}$ for the existence of the $\rm H_2 O$ maser
    (Cao \& Jiang 1997). It can be clearly found that the estimated accretion rate in maser disk
    is much lower than that constrained from the inner RIAF, where $\dot{M_{\rm tr}}\sim 10^{-2}\msun/\rm yr$.
     We note that this is possible since that the hot gas inside the Bondi radius will also be accreted beside
    the accreting cold gas through the maser disk, where the Bondi radius, $R_{\rm B}\sim 10^6 R_{\rm g}$,
    is larger than the outer radius of the maser disk. The high-resolution X-ray observation on the diffused
    gas near the Bondi radius will help to test this issue. The accretion rate of the outer thin disk may
    be not a constant considering the mixture of inflowing hot and cold gas, where the hot gas will be
    condensed down to the SSD gradually due to the cooling. Actually, the estimated accretion rate in
    the inner edge of maser disk (Cao \& Jiang 1997) is higher than that in outer edge (Neufeld \& Maloney 1995;
    Herrnstein et al. 2005), which may more or less support this scenario. In calculating the spectrum of
    outer SSD, we simply use the rate at transition radius, which will not affect our main conclusion
    since that most gravitational energy of the thin disk is released in its inner region.

   The warp of outer maser disk could has various causes. Perhaps, the most promising mechanism should be
   the Bardeen-Petterson effect that caused by the interaction of the accretion disk and the spinning BH
   if their spin axes are misaligned (see Introduction). The superposed maps of maser emission in NGC 4258
   for all epochs suggest that the warp radius should be around 4 mas ($\sim8\times 10^4 R_{\rm g}$), where
   the innermost maser spots at $\sim$ 3 mas deviate from the others evidently (see Figure 6 in Argon et al. 2007).
   Our estimated warp radius, $R_{\rm w}\simeq8.6\times10^4 R_{\rm g}$, based on the Bardeen-Petterson effect
   (equation 2), is consistent with the observational results remarkably well. This result support that the
   Bardeen-Petterson effect may be the physical reason for the warped disk in NGC 4258. Caproni et al. (2007)
   also estimated the warp radius of NGC 4258 by comparing the timescales of Lense-Thirring precession and
   the warp transmission through the disk based on a parametric disk model. They found that the warp radius
   is comparable to or smaller than the radius of the inner maser, which is more or less consistent with our
   result. Due to the uncertainties of the parametric disk model and the poorly known viscosities in azimuthal
   and vertical direction of the disk, it prevent Caproni et al. (2007) to give a strong constraint on the BH
   spin from observations.

\section*{ACKNOWLEDGMENTS}
  We thank the anonymous referee for useful and constructive comments and suggestions. We also thank Ju-Fu Lu,
  Xinwu Cao and members of HUST astrophysics group for many useful discussions and comments.
  We also grateful to Feng Yuan for his comments/suggestions and for allowing us using his code of jet emission.
   This work was supported by the National Basic Research Program of China (2009CB824800), the NSFC (grants 11103003 and 11133005).


\begin{thebibliography}{}


\bibitem[Argon et al.(2007)]{ar07} Argon, A.L., Greenhill, L. J., Reid, M. J., Moran, J. M., \& Humphreys, E. M. L. 2007, ApJ, 659, 1040

\bibitem[Bardeen \& Petterson(1975)]{bp75} Bardeen, J. M., \& Petterson, J. A. 1975, ApJ, 195, 65


\bibitem[Blandford \& Znajek(1977)]{bz77} Blandford, R. D. \& Znajek, R. L. 1977, MNRAS, 179, 433

\bibitem[Blandford \& Begelman(1999)]{bb99} Blandford, R. D., \& Begelman, M. C. 1999, MNRAS, 303, L1

\bibitem[Birzan(2008)]{bi08} Birzan, L., McNamara, B., Nulsen, P., Carilli, C., \& Wise, M. 2008, ApJ, 686, 859

\bibitem[Cao \& Spruit(2013)]{ch13}  Cao, X. \& Spruit, H. C. 2013, ApJ, 765, 149

\bibitem[Cao(2012)]{cao12}  Cao, X. 2012, MNRAS, 426, 2813

\bibitem[Cao \& Rawlings(2004)]{ca04} Cao, X., \& Rawlings, S. 2004, MNRAS, 349, 1419

\bibitem[Cao \& Jiang(1997)]{ca97} Cao, X., \& Jiang, D. R. 1997, A\&A, 322, 49

\bibitem[Caproni et al.(2007)]{ca07} Caproni, A., Abraham, Z., Livio, M., \& Mosquera Cuesta, H. J. 2007, MNRAS, 379, 135

\bibitem[Cavagnolo et al.(2010)]{ca10} Cavagnolo, K. W., et al. 2010, ApJ, 720, 1066

\bibitem[Cecil et al.(2000)]{ce00} Cecil, G., et al. 2000, ApJ, 536, 675

\bibitem[Chary et al.(2000)]{ch00} Chary, R., Becklin, E. E., Evans, A. S., Neugebauer, G., Scoville, N. Z., Matthews, K., \& Ressler, M. E. 2000, ApJ, 531, 756

\bibitem[De Villiers et al.(2005)]{de05} De Villiers, J.-P., et al. 2005, ApJ, 620, 878

\bibitem[Di Matteo et al.(2003)]{di03} Di Matteo, T., Allen, S. W., Fabian, A. C., Wilson, A. S., \& Young, A. J. 2003, ApJ, 582, 133

\bibitem[Doi et al.(2013)]{do13} Doi, A., Asada, K., Fujisawa, K., Nagai, H., Hagiwara, Y., Wajima, K., \& Inoue, M. 2013, ApJ, 765, 63

\bibitem[Falcke \& Biermann(1999)]{fb99} Falcke, H., \& Biermann, P. L. 1999, A\&A, 342, 49

\bibitem[Fathi et al.(2011)]{fa11} Fathi, K., Axon, D. J., Storchi-Bergmann, T., et al. 2011, ApJ, 736, 77

\bibitem[Freedman et al.(2001)]{fr01} Freedman, W. L., et al. 2001, ApJ, 553, 47

\bibitem[Fruscione et al.(2005)]{fr05} Fruscione, A., Greenhill, L. J., Filippenko, A. V., Moran, J. M., Herrnstein, J. R., \& Galle, E. 2005, ApJ, 624, 103

\bibitem[Gammie et al.(1999)]{ga99} Gammie, C. F., Narayan, R., Blandford, R. 1999, ApJ, 516, 177

\bibitem[Giroletti et al.(2005)]{gi05} Giroletti, M., Taylor, G. B., \& Giovannini, G. 2005, ApJ, 622, 178

\bibitem[Gou et al.(2011)]{go11} Gou, Lijun, et al. 2011, ApJ, 742, 85

\bibitem[Hales et al.(1988)]{ha88} Hales, S. E. G., Baldwin, J. E., Warner, P. J. 1988, MNRAS, 234, 919

\bibitem[Hawley \& Krolik(2006)]{hk06} Hawley, J. F., \& Krolik, J. H. 2006, ApJ, 641, 103

\bibitem[Hawley et al.(2006)]{ha96} Hawley, J.F., Gammie, C.F., \& Balbus, S.A., 1996, ApJ, 464, 690.

\bibitem[Herrnstein et al.(1996)]{he96} Herrnstein, J. R., Greenhill, L. J. \& Moran, J. M. 1996, ApJ, 468, 17

\bibitem[Herrnstein et al.(1997)]{he97} Herrnstein, J. R., Moran, J. M., Greenhill, L. J., Diamond, P. J., Miyoshi, M., Nakai, N., \& Inoue, M. 1997, ApJ, 475, 17

\bibitem[Herrnstein et al.(1998)]{he98} Herrnstein, J. R., Greenhill, L. J., Moran, J. M., Diamond, P. J., Inoue, M., Nakai, N., Miyoshi, M. 1998, ApJ, 497, 69

\bibitem[Herrnstein et al.(1999)]{he99} Herrnstein, J. R., et al. 1999, Nature, 400, 539

\bibitem[Herrnstein et al.(2005)]{he05} Herrnstein, J. R., Moran, J. M., Greenhill, L. J., Trotter, Adam S. 2005, ApJ, 629, 719

\bibitem[Hirose et al.(2004)]{hi04} Hirose, S., et al. 2004, ApJ, 606, 1083

\bibitem[Ho(2008)]{ho08} Ho, L. C. 2008, ARA\&A, 46, 475

\bibitem[Ho \& Peng(2001)]{hp01} Ho, L. C., \& Peng, C. Y. 2001, ApJ, 555, 650

\bibitem[Ho \& Ulvestad(2001)]{hu01} Ho, L. C., \& Ulvestad, J. S. 2001, ApJS, 133, 77

\bibitem[Ho et al.(1997)]{ho97} Ho, L. C., Filippenko, A. V., \& Sargent, W. L. 1997, ApJ, 112, 315

\bibitem[Ichimaru(1977)]{ic77} Ichimaru, S. 1977, ApJ, 214, 840

\bibitem[Krause et al.(2007)]{kr07} Krause, M., Fendt, C., \& Neininger, N. 2007, A\&A, 467, 1037

\bibitem[Lasota et al.(1996)]{la96} Lasota, J.-P., Abramowicz, M. A., Chen, X., Krolik, J., Narayan, R., Yi, I. 1996, ApJ, 462, 142

\bibitem[Laine et al.(2010)]{la10} Laine, S., Krause, M., Tabatabaei, F. S., \& Siopis, C. 2010, AJ, 140, 1084

\bibitem[Lei et al.(2013)]{le13} Lei, W. -H., Zhang, B., \& Gao, H. 2013, ApJ, 762, 98

\bibitem[Li \& Cao (2009)]{lc09} Li, S.-L., Cao, X. 2009, MNRAS, 400, 1734

\bibitem[Liu \& Wu(2013)]{liu13} Liu, H., \& Wu, Q. 2013, ApJ, 764, 17

\bibitem[Liu, et al.(2010)]{liu10}  Liu, T., Liang, E.-W., Gu, W.-M., Zhao, X.-H., Dai, Z.-G., \& Lu, J.-F. 2010, A\&A, 516, 16

\bibitem[Lu \& Zhou(2005)]{lu05} Lu, J.-F., \& Zhou, B.-Y. 2005, ApJL, 635, 17

\bibitem[Lu et al.(2004)]{lu04}  Lu, J.-F., Lin, Y.-Q., Gu, W.-M. 2004, ApJ, 602, 37

\bibitem[Manmoto(2000)]{man00} Manmoto, T. 2000, ApJ, 534, 734

\bibitem[Maloney et al.(1996)]{ma96} Maloney, P. R., Begelman, M. C., \& Pringle, J. E. 1996, ApJ, 472, 582

\bibitem[Markoff et al.(2008)]{mak08} Markoff, S., et al. 2008, ApJ, 681, 905

\bibitem[Markoff et al.(2001)]{ma01} Markoff, S., Falcke, H., \& Fender, R. 2001, A\&A, 372, 25

\bibitem[Martin(2008)]{ma08} Martin, R. G. 2008, MNRAS, 387, 830

\bibitem[Martinez-Sansigre \& Rawlings(2011)]{mr11} Martinez-Sansigre, A., \&  Rawlings, S. 2011, MNRAS, 414, 1937

\bibitem[Mason et al.(2012)]{ma12} Mason, R. E., et al. 2012, AJ, 144, 11

\bibitem[McKinney et al.(2012)]{mc12} McKinney, J. C., Tchekhovskoy, A., \& Blandford, R. D. 2012, MNRAS, 423, 3083

\bibitem[McKinney \& Gammie(2004)]{mg04} McKinney, J. C. \& Gammie, C. F. 2004, ApJ, 611, 977

\bibitem[McKinney(2005)]{mc05} McKinney, J. C. 2005, ApJ, 630, 5

\bibitem[Medvedev(2006)]{me06} Medvedev, M. V. 2006, ApJ, 651, 9

\bibitem[Meier(2001)]{me01}  Meier, D. L. 2001, ApJ, 548, L9

\bibitem[Miyoshi et al.(1995)]{mi95} Miyoshi, M., Moran, J., Herrnstein, J., Greenhill, L., Nakai, N., Diamond, P., \& Inoue, M. 1995, Nature, 373, 127

\bibitem[Moran et al.(1995)]{mo95} Moran, J., Greenhill, L., Herrnstein, J., Diamond, P., Miyoshi, M., Nakai, N., \& Inque, M., 1995, PNAS, 92, 11427


\bibitem[Nagar et al.(2001)]{na01} Nagar, N. M., Wilson, A. S., \& Falcke, H. 2001, A\&A, 559, 87

\bibitem[Narayan et al.(2012)]{na12} Narayan, R., Sadowski, A., Penna, R. F., Kulkarni, A. K. 2012, MNRAS, 426, 3241

\bibitem[Narayan \& McClintock(2012)]{nm12} Narayan, R. \& McClintock, J. E. 2012, MNRAS, 419, 69

\bibitem[Narayan \& McClintock(2008)]{nm08} Narayan, R. \& McClintock, J. E. 2008, New Astronomy Reviews, 51, 733

\bibitem[Narayan \& Yi(1995)]{ny95} Narayan, R., \& Yi, I. 1995, ApJ, 452, 710

\bibitem[Narayan \& Yi(1994)]{ny94} Narayan, R., \& Yi, I. 1994, ApJ, 428, 13

\bibitem[Nemmen et al.(2012)]{ne12} Nemmen, R., Storchi-Bergmann, T., \& Eracleous, M. 2012, submitted to ApJ, arXiv:1112.4640

\bibitem[Nemmen et al.(2007)]{ne07} Nemmen, R. S., Bower, R. G., Babul, A., \& Storchi-Bergmann, T. 2007, MNRAS, 377, 1652

\bibitem[Nemmen et al.(2006)]{ne06} Nemmen, R. S., Storchi-Bergmann, T., Yuan, F., Eracleous, M., Terashima, Y., \& Wilson, A. S. 2006, ApJ, 643, 652

\bibitem[Neufeld \& Maloney(1995)]{ne95} Neufeld, D. A., \& Maloney, P. R. 1995, ApJL, 447, 17

\bibitem[Papaloizou et al.(1998)]{pa98} Papaloizou, J. C. B., Terquem, C., \& Lin, D. N. C. 1998, ApJ, 497, 212

\bibitem[Piran(1999)]{pi99} Piran, T. 1999, Physics Reports, 314, 575

\bibitem[Pringle(1996)]{pr96} Pringle, J. E. 1996, MNRAS, 281, 357

\bibitem[Quataert et al.(1999)]{qu99} Quataert, E., Di Matteo, T., Narayan, R., \&  Ho, L. C. 1999, ApJ, 525, 89

\bibitem[Reid(1993)]{re93} Reid, M. J. 1993, ARA\&A, 31, 345

\bibitem[Reynolds et al.(2009)]{re09} Reynolds, C. S., Nowak, M. A., Markoff, S., Tueller, J., Wilms, J., Young, A. J. 2009, ApJ, 691, 1159

\bibitem[Sarazin et al.(1980)]{sa80} Sarazin, C. L., Begelman, M. C., \& Hatchett, S. P. 1980, ApJ, 238, L129

\bibitem[Shu et al.(2011)]{sh11}  Shu, X. W., Yaqoob, T., \& Wang, J. X. 2011, ApJ, 738, 147

\bibitem[van der Kruit et al.(1972)]{van72} van der Kruit, P. C., Oort, J. H., \& Mathewson, D. S. 1972, A\&A, 21, 169

\bibitem[Wilkes et al.(1995)]{wi95} Wilkes, B. J., Schmidt, G. D., Smith, P. S.; Mathur, S., McLeod, K. K. 1995, ApJ, 455, 13

\bibitem[Wu et al.(2011)]{wu11} Wu, Q., Cao, X., \& Wang, D. X. 2011, ApJ, 735, 50

\bibitem[Wu \& Cao(2008)]{wu08} Wu, Q., \& Cao, X. 2008, ApJ, 687, 156

\bibitem[Wu et al.(2007)]{wu07} Wu, Q., Yuan,F.,\& Cao, X. 2007, ApJ, 669, 96

\bibitem[Xu \& Cao(2009)]{xc09} Xu, Y.-D. \& Cao, X 2009, RAA, 9, 401


\bibitem[Yu et al.(2011)]{yyh11} Yu, Z., Yuan, F., \& Ho, L. C. 2011, ApJ, 726, 87

\bibitem[Yuan et al.(2012)]{yu12}  Yuan, F., Wu, M., Bu, D. 2012, ApJ, 761, 129

\bibitem[Yuan et al.(2009)]{yu09} Yuan, F., Yu, Z., \& Ho, L. C. 2009, ApJ, 703, 1034

\bibitem[Yuan et al.(2005)]{ycn05} Yuan, F., Cui, W., \& Narayan, R. 2005, ApJ, 620, 905

\bibitem[Yuan \& Cui(2005)]{yc05} Yuan, F., \& Cui, W.,  2005, ApJ, 629, 408

\bibitem[Yuan et al.(2003)]{yu03} Yuan, F., Quataert, E., \& Narayan, R. 2003, ApJ, 598, 301

\bibitem[Yuan et al.(2002)]{yu02} Yuan, F., Markoff, S., Falcke, H., \& Biermann, P. L. 2002, A\&A, 391, 139

\bibitem[Zhang et al.(1997)]{zh97} Zhang, S. N., Cui, W., \& Chen, W. 1997, ApJ, 482, L155

\bibitem[Shakura \& Sunyaev(1973)]{ss73} Shakura, N. I., \& Sunyaev, R. A. 1973, A\&A, 24, 337

\bibitem[Sharma et al.(2007)]{sh07} Sharma, P., Quataert, E., Hammett, G. W., \& Stone, J. M. 2007, ApJ, 667, 714

\bibitem[Steiner et al.(2013)]{st13} Steiner, J. F., McClintock, J. E., \& Narayan, R. 2013, ApJ, 762, 104

\bibitem[Turner \& Ho(1994)]{th94} Turner, J. L., \& Ho, P. T. P. 1994, ApJ, 421, 122

\bibitem[Tchekhovskoy et al.(2011)]{tc11} Tchekhovskoy, A., Narayan, R., \& McKinney, J. C. 2011, MNRAS, 418, 79

\bibitem[Tchekhovskoy et al.(2010)]{tc10} Tchekhovskoy, A., Narayan, R., McKinney, J. C. 2010, ApJ, 711, 50

\end{thebibliography}
\end{document}